\documentclass{article}

\usepackage{PRIMEarxiv}
\usepackage[utf8]{inputenc} 
\usepackage[T1]{fontenc}    
\usepackage{hyperref}       
\usepackage{url}            
\usepackage{booktabs}       
\usepackage{amsfonts}       
\usepackage{nicefrac}       
\usepackage{microtype}      
\usepackage{lipsum}
\usepackage{fancyhdr}       
\usepackage{graphicx}       
\graphicspath{{media/}}
\usepackage{graphicx} 
\usepackage{float} 
\usepackage[
    backend=biber,
    style=numeric,
    sorting=none
  ]{biblatex}

\usepackage{biblatex} 
\title{Entendre, a Social Bot Detection Tool for Niche, Fringe, and Extreme Social Media}

\author{
  Pranav Venkatesh \\
  Computational Media Lab \\
  Department of Computer Science \\
  The University of Texas at Austin \\
  \texttt{pranav.vankatesh@utexas.edu} \\
  \And
  Kami Vinton \\
  Computational Media Lab \\
    School of Journalism and Media \\
  The University of Texas at Austin \\
  \texttt{kamivinton@utexas.edu} \\
  \And
  Dhiraj Murthy \\
  Computational Media Lab \\
  School of Journalism and Media \\
  Moody College of Communication \\
  The University of Texas at Austin \\
  \texttt{dhiraj.murthy@austin.utexas.edu} \\
  \And
    Kellen Sharp\\
  Computational Media Lab \\
Department of Radio-Television-Film\\
  The University of Texas at Austin \\
  \texttt{knsharp@utexas.edu} \\
  \And
  Akaash Kolluri \\
  Computational Media Lab \\
  The University of Texas at Austin \\
  \texttt{akaashrkolluri@computationalmedialab.com} \\
  \\
}

\begin{document}
\maketitle

\begin{abstract}
Social bots—automated accounts that generate and spread content on social media—are exploiting
vulnerabilities in these platforms to manipulate public perception and disseminate disinformation.
This has prompted the development of public bot detection services; however, most of these services
focus primarily on Twitter, leaving niche platforms vulnerable. Fringe social media platforms such as
Parler, Gab, and Gettr often have minimal moderation, which facilitates the spread of hate speech
and misinformation. To address this gap, we introduce Entendre, an open-access, scalable, and
platform-agnostic bot detection framework. Entendre can process a labeled dataset from any social
platform to produce a tailored bot detection model using a random forest classification approach,
ensuring robust social bot detection. We exploit the idea that most social platforms share a generic
template, where users can post content, approve content, and provide a bio (common data features).
By emphasizing general data features over platform-specific ones, Entendre offers rapid extensibility
at the expense of some accuracy. To demonstrate Entendre’s effectiveness, we used it to explore the
presence of bots among accounts posting racist content on the now-defunct right-wing platform Parler.
We examined 233,000 posts from 38,379 unique users and found that 1,916 unique users (4.99\%)
exhibited bot-like behavior. Visualization techniques further revealed that these bots significantly
impacted the network, amplifying influential rhetoric and hashtags (e.g., \#qanon, \#trump, \#antilgbt).
These preliminary findings underscore the need for tools like Entendre to monitor and assess bot
activity across diverse platforms.
\end{abstract}

\keywords{Entendre \and Social Bots \and Bot Detection \and Machine Learning \and Botometer \and Crowdsourcing \and Parler \and Metadata \and Sequential Model-Based Optimization \and Bot Interactions}

\section{Introduction}
A social bot is a computer algorithm designed to automatically generate and share content on social media platforms,
engaging with human users and mimicking their behavior with the potential to modify it \cite{Ferrara2016}. Social bots cause a wide array of adverse effects on social media platforms. Specifically, they exploit platform vulnerabilities, which can include stealing user data, large-scale surveillance, and spreading malware and misinformation \cite{BOSHMAF2013}. Additionally, social bots can spread disinformation (e.g.,   to further the agendas of  partisan agendas) \cite{Choi2020}. These bots can  influence public perceptions and falsely inflate perceived support for radical, fringe agendas that drown out mainstream voices in political debates and increase the marginalization of non-dominant groups \cite{Choi2020}, introduce variability to algorithmic trading \cite{Nam2022}, and control botnets. When leveraged, these bots
can influence public perception and artificially inflate perceived support for radical, fringe agendas, which in turn drown
out mainstream voices in political debates and increase the marginalization of non-dominant groups \cite{Choi2020}. Governments and political actors have previously used social bots to successfully manipulate public opinion \cite{Marcellino2020}. Malicious social bots  have been responsible for amplifying and strengthening campaigns that disseminate hateful content, manipulate users, and instigate real-world violence \cite{Wahlström2021,Gruzd_Mai}. Unfortunately, social bots can be difficult to detect, and often outmaneuver mitigation algorithms \cite{Marcellino2020}. Likewise, there are few, if any, open-source bot detection tools available for fringe social media platforms. Despite the documented, outsized societal harms these “niche” platforms facilitate, there remains a paucity of computational tools designed specifically to identify, intercept, and remove or mitigate known causes, such as social bots \cite{Choi2020, Wahlström2021, Orabi2020}. As mainstream platforms become more proficient at identifying and removing malicious social bots, they will undoubtedly become even more prolific on fringe social media platforms.

The malicious use of social bots prompted the creation of public bot detection services. However, their overwhelming focus has been on Twitter \cite{Orabi2020}. We help address this imbalance and  broaden bot detection approaches by employing our tool on Parler, a US right wing social media platform. In this paper, we demonstrate its  efficacy, while  providing flexibility and extensibility  for it to be quickly adapted to other platforms. Public bot detection services democratize the detection, monitoring, and analysis of bot activity, and help to intercept and mitigate some of their harmful effects. In addition to an abundant focus on Twitter, bot-detection tools tend to be platform-specific (e.g., Botometor), and lack flexibility to easily adapt them for use on other programs. Likewise, academic and industry developers have neglected creating these tools for use on smaller, niche, and extreme social media platforms (e.g., Parler, Gab, and Gettr). The story of Parler’s sudden rise in popularity and subsequent involvement in offline violence is a salient example of the urgent need to address this gap. In 2020, Parler's user base grew exponentially over a short period of time once Twitter began actively enforcing its terms of service (e.g., deleted or labeled misleading posts by Donald Trump and many other conservatives \cite{Leskin2019}). Parler was billed as a Twitter alternative \cite{Baines}. With no meaningful moderation, the platform soon became a hate-filled, misinformative, echo chamber fomenting further division in the American electorate and eventually led to coordinated, offline violence \cite{Aliapoulios2021, Hitkul2021}. Parler was eventually de-platformed by Amazon for its role in coordinating and fomenting violent acts during the January 6th, U.S. Capitol Insurrection \cite{Aliapoulios2021}. The platform eventually resurfaced under new registration and, after being subjected to stricter content moderation, was
relaunched in the Apple App Store on May 17, 2021 \cite{Molina2021} and soon afterward on Google Play on September
2, 2022 \cite{Grant2022}. On April 14, 2023, the media conglomerate Starboard acquired Parler and promptly shut it down
the same day. This shut down, however, may be temporary as the new owners optimistically hint at a 2024 relaunch,
citing the precedent set by Apple and Google’s hosting of X (formerly Twitter), despite the platform’s increasingly
"lenient" speech policies \cite{IngramBrooks2024}. This situation underscores the necessity of continuous development of
tools.


To  help address the dearth of bot detection tools for extreme, fringe social media platforms, we developed Entendre, an open-access bot detection framework for Parler. Entendre checks the activity of requested accounts on Parler and utilizes feature-based machine learning to predict the percentage likelihood of the account being a bot via a public facing web application and API. Entendre can also be easily extended to other niche social media platforms such as Gab and Gettr. Our framework aims to: 1) Provide a scalable, adaptable social bot detection tool  optimized for Parler and 2) Highlight the urgent need for bot detection on more extreme platforms to better generalize whether humans or automated actors are responsible for particular types of content dissemination. Given the complexities surrounding disinformation and their sources, Entendre extends and develops previous work \cite{Ferrara2016} by specifically focusing on Parler.

\section{Previous Work}
Botometer, is an example of a successful, publicly-available service that classifies the extent to which accounts on Twitter exhibit bot-like behavior \cite{Ferrara2016}. \cite{Gruzd_Mai} employed Botometer to determine the concentration of automated accounts behind the \#FilmYourHospital COVID-19 misinformation campaign on Twitter. Likewise, Botometer was used to identify the 2016 Russian interference campaign, finding that 4.9\% of liberals and 6.2\% of conservatives were social bots, and they significantly amplified Russian trolls’ content. Indeed, a web application using Botometer to visualize levels of bot activity and their targeted topics \cite{Davis2016}. The impact of bot detection, along with the adversarial race between the development of
bot detection tooling and advanced bots, has been emphasized \cite{Yang2022}.

One recent study demonstrated compelling (albeit indirect) evidence that social bots may be rampant on Parler, possibly having an outsized influence exacerbating the echo chamber effects previously observed on the platform \cite{Hitkul2021}. For example, the top five users on Parler accounted for 11\% of all content \cite{Hitkul2021}, which underscores the urgent need to directly measure the presence, activities, and magnitude of social bots on Parler and other largely unmoderated, fringe social media platforms.

\section{Methods}
\subsection{Dataset}
To train our machine learning model, we used an open dataset scraped from Parler, containing 183 million posts made by 4 million users between August 2018 and January 2021 \cite{Aliapoulios2021}. We constructed a data subset of verified social bot and human accounts. Each post and account contained relevant metadata (followers, hashtags, links, etc). To build the subset, human annotation and automated techniques (flagging suspicious accounts for human review) labeled a significant amount of social bot and human accounts, yielding 50,000 posts for model training. For the pre-processing step, we normalized features to reduce the training time and ensure data consistency. For missing features, we imputed the missing information through a manual scrape of Parler. For model feature selection, we followed the methodology outlined by \cite{Schnebly} that focuses on choosing basic account attributes that delineate account usage and deriving nuanced attributes of the account.

\subsection{Machine Learning Model}
In general, there are three approaches for social bot detection: 1) graph-based, 2) crowdsourcing, and 3) feature-based machine learning \cite{Ferrara2016}. We opted for a feature-based machine learning approach due to its rapid classification when given new data points. Our machine learning algorithm classifies social bot and human accounts based on recent posts and metadata. Extensive literature has pointed to the use of the random forest classification model for robust social bot detection over other models due to its effective classification of numerical inputs \cite{Schnebly}, performance accuracy, and wide popularity for social bot detection \cite{Alothali}. For hyperparameter tuning of the random forest model (node size, number of trees, etc), we followed \cite{Probst2019} detailed strategy of sequential model-based optimization to improve the total performance of our model.

\begin{figure}[h]
    \centering
    \includegraphics[width=0.55\textwidth]{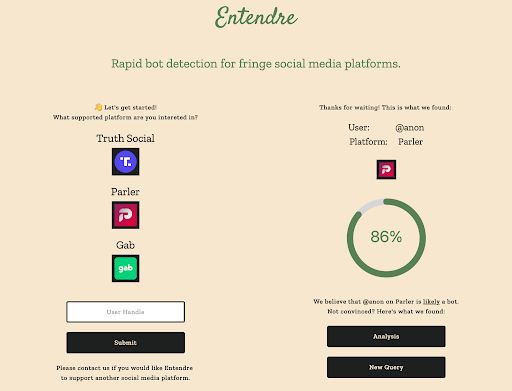}
    \caption{Frontend of Entendre.}
    \label{fig:entendre_1}
\end{figure}

\subsection{Backend API}
Entendre's backend API was written with Express.js. The API, after being provided with a Parler account, will return the percentage likelihood of that account being a bot. While Parler does not have an official API, the platform had numerous security exploits \cite{Isbitski2024} and it remains to be seen whether the new, post-de-platformed incarnation of Parler is more secure or not. Entendre uses an unofficial Parler API \cite{Iturbe_API} to retrieve requested account posts along with relevant metadata. The collected information feeds into a supervised machine learning model which computes and returns a bot-likelihood score for the account. The backend API is publicly facing so it can be used directly for bot detection.

\subsection{Frontend Website}
Entendre is additionally available through a public-facing web application created in React.js (see Figure 1). Like
Botometer \cite{Ferrara2016} , a request initiated from the web application similarly returns the percentage likelihood of an account being a bot and generates relevant visuals for further analysis, specifically time-based graphs of account
activity and charts of most frequently used terms/hashtags.

\section{Preliminary Findings}
We employed a prototype version of Entendre over a sample of the open Parler dataset \cite{Aliapoulios2021} to understand the nature of bot interactions and presence on the Parler platform. The prototype version of Entendre exists as a local Python toolchain, allowing for a direct interface with our core bot prediction software to enable for rapid classification large-scale classification. The prototype version of Entendre was built prior to obtaining a labeled dataset of bot posts to train the machine learning model. For preliminary testing, the prototype version of Entendre utilizes a mix of fuzzy matching (to identify spam content across account posts) and basic heuristics reflective of social bot posts (high volume of posts:  > 100 posts/day, follower to following ratio: suspiciously close to 1, indicating a network of bots, etc) to predict the bot likelihood of a given user. 
Given the breadth and impact of hateful content on Parler \cite{Aliapoulios2021}, we utilized Entendre to explore the presence of bots among accounts that post racist content. We derived a set of racist users by using a seed set of racist posts and expanding the set by adding content connected to it, yielding a working dataset of 233K posts from 38379 unique users. By feeding the metadata attached with each individual user paired along with their respective posts in the sample of Parler into Entendre, we were able to classify 1916 unique users as displaying bot-like activity (4.99\% of unique users in the dataset). 

\begin{figure}[h]
    \centering
    \includegraphics[width=0.8\textwidth]{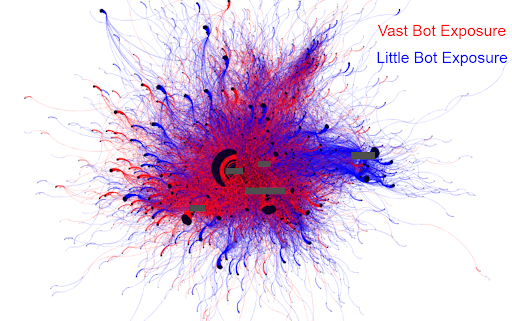}
    \caption{Sample microcosm network visual of bot activity within the Parler ecosystem wherein nodes represent unique user accounts and edges represent user engagement. Red nodes indicate vast bot exposure, while blue nodes indicate little bot exposure.}
    \label{fig:entendre_2}
\end{figure}

We visualized a sample macrocosm (see Figure 2) via network based methods using Gephi \cite{Bastian2009}, where unique users represent nodes and directed edges from the original poster to replier represent post engagement (mentions, replies, comments). The nodes were colored red if the user either displayed bot-like activity or engaged with a majority bot content and blue if the user engaged with a majority non-bot content. The edges were colored according to the color classification of the head node to better emphasize the impact of automated content. The network visualization was clustered using the ForceAtlas2 algorithm.

Figure 2 illustrates how bots on Parler propagate content across the platform. The bot accounts clearly coalesce, forming a massive red colored diffusion of automated content to other  users across the Parler ecosystem. Further analysis was conducted on a user-level, analyzing the characteristics of high-impact automated accounts. For example, one user account, which was found to have considerable influence on the network (based on eigenvector centrality) often made posts echoing the rhetoric of popular influencers on the platform followed by a series of hashtags (\#qanon, \#trump, \#antilgbt, \#christian, and \#wwg1wga, amongst other hashtags). The visualization paired with the content being excreted by these automated accounts implies the bot’s instigator-like role in the Parler ecosystem, spreading the posts of popular influencers across the network as a whole. This  further underscores the urgent necessity of tools like Entendre. By providing a scalable method for detection of bots, alternative, fringe social media platforms like Parler can more easily identify bots and take action to reduce their impact.

\section{Conclusion}
In this paper, we introduced Entendre, a robust framework for the detection of bots on niche, fringe, and extreme social media platforms. Utilizing a prototype version of Entendre, we have shown the nature and presence of bots in instigating and propagating harmful content on alternative platforms, demonstrating the urgent need for accurate and scalable bot detection. 

\section{Acknowledgments}
This was supported in part by Rishul Bhuvanagiri for his contributions in LaTeX coding with references and formatting. This work was supported by Good Systems, a research grand challenge at the University of Texas at Austin.

\printbibliography 

\end{document}